\documentclass[
	fontsize=11pt,         
	numbers=noenddot,  
    parskip=half,        	
    listof=totoc,        	
	headsepline=true,      
	footsepline=false, 	
    DIV=12          
]{scrartcl}

\usepackage[utf8]{inputenc}
\usepackage[T1]{fontenc}
\usepackage[english]{babel}

\usepackage{graphicx}
\usepackage{geometry}
\usepackage{nicefrac}
\usepackage{csquotes}
\usepackage{subcaption}
\usepackage{longtable}
\usepackage{xcolor}
\usepackage{booktabs}
\usepackage{multirow}
\usepackage{microtype}
\usepackage{natbib}
\usepackage{lmodern}

\usepackage{amsmath, amsfonts, amssymb, bbm, bm}
\usepackage{pdflscape}

\usepackage[onehalfspacing]{setspace}

\usepackage{scrlayer-scrpage}
\clearscrheadings                   
\automark{section}                  
\ihead{}			
\ohead{Knop, Pauly, Friede \& Welz: Why interactions matter}				
\ifoot{}							
\ofoot{}

\usepackage{booktabs}

\usepackage[hidelinks]{hyperref}

\begin{document}

\thispagestyle{empty}

\vspace{2em}

\begin{center}
\LARGE{\textbf{The impact of neglected confounding and interactions in mixed-effects meta-regression}\\[1cm]
}
\end{center}

\begin{center}
\large{Eric S. Knop\footnote{
Department of Statistics, TU Dortmund University}, Markus Pauly\footnotemark[1]\footnote{UA Ruhr, Research Center Trustworthy Data Science and Security, Dortmund}, Tim Friede\footnote{Department of Medical Statistics, University of Göttingen}\footnote{DZHK (German Center for Cardiovascular Research), partner site Göttingen} and Thilo Welz\footnotemark[1]}
\end{center}

\begin{center}
 \today  
\end{center}

\begin{small}
    \textbf{Abstract}
Analysts seldom include interaction terms in meta-regression model, what can introduce bias if an interaction is present. We illustrate this in the current paper by re-analyzing an example from research on acute heart failure, where neglecting an interaction might have led to erroneous inference and conclusions. Moreover, we perform a brief simulation study based on this example highlighting the effects caused by omitting or unnecessarily including interaction terms. Based on our results, we recommend to  always include interaction terms in mixed-effects meta-regression models, when such interactions are plausible.
\end{small}


\section{Introduction}
In meta-regression, interactions reflecting effect moderators are often neglected. Certainly, in many situations, ignoring interactions is due to the small number of available studies \citep{davey2011}. But even if enough studies are available, interactions are often not considered (e.g. \cite{andreas2019}; \cite{kimmoun2021}). As we know from ordinary least squares regression, this can lead to biased estimates, whereas including redundant moderators and interactions only increases the variance of the estimates (\cite{Greene2019}, pp. 99-101).\\
In our work, we illustrate these points by a reanalysis of a recent meta-analysis, including mixed-effects meta-regression analyses \citep{kimmoun2021}. \cite{kimmoun2021} analyzed mortality and readmission to hospital after acute heart failure in clinical studies. They found a statistically significant decline of mortality over calendar time. 
However, the average age of the patients also decreased over calendar time. This suggests that the observed trend might at least partially be explained by a neglected interaction between the average age and the median recruitment year. Therefore, we conducted a meta-regression where we included the median recruitment year, the average age and their interaction. Additionally, we conducted a simulation study mimicking the data example to assess the impact of model misspecification. In the simulation, we either assumed the model with only recruitment year as moderator or the model with both moderators and their interaction to be the true model. In both scenarios we fitted a model with (i) one moderator, (ii) two moderators or (iii) two moderators and their interaction. Confidence intervals for the moderators were compared regarding their empirical coverage probabilities and median interval lengths as well as empirical biases of the coefficient estimators.

\section{Methods}
A mixed-effects meta-regression model with two moderators and an interaction term is of the form 
\begin{equation}
y_i = \beta_0 + x_{1i}\beta_1+x_{2i}\beta_2 + x_{1i}x_{2i}\beta_{12} + u_i + e_i, i=1, ..., k
\label{eq:om}    
\end{equation} 
where $k$ denotes the number of studies, $y_i$ is a function of the effect measure, $x_{1i}$ and $x_{2i}$ denote the moderators, $u_i$ the random effect and $e_i$ the sampling error of the study $i=1, ..., k$. The $u_i$ and $e_i$ are usually assumed to be independent and follow normal distributions with  $u_i\sim \mathcal{N}(0, \tau^2)$ and $e_i\sim\mathcal{N}(0, v_i)$ \citep{raudenbush2009}. In the following analysis, the parameter vector $\boldsymbol{\beta}=(\beta_0, \beta_1, \beta_2, \beta_{12})^T$ is estimated via weighted least squares regression using inverse variance weights \citep{Berkey1995}. The variance of the random effect $\tau^2$ is estimated via restricted maximum likelihood estimation (\cite{fahrmeir2013}, pp. 372-373) since it has been recommended as a good
choice of estimator for continuous outcomes in other research \citep{veroniki2016}. The confidence intervals for the moderators are calculated using the Knapp-Hartung method \citep{knapphartung2003}, because it was shown to be more accurate compared to other covariance estimators in simulation studies (\cite{viechtbauer2015}; \cite{welzpauly2020}). The models with only one and only two moderators in Sections \ref{sec:meta_reg} and \ref{sec:sim} can be seen as special cases of the model in equation (\ref{eq:om}). All analyses were conducted in \texttt{R 4.1.1} \citep{r} using the \texttt{metafor} package \citep{metafor}.The R-Code will be made publicly available. \\

\section{Meta Regressions for Acute Heart Failure Studies}\label{sec:meta_reg}
The research synthesis by \cite{kimmoun2021} included 285 studies on acute heart failure, published between 1998 and 2017. Outcome measures were $30$-day and $1$-year readmission rates and $30$-day and $1$-year mortality in 108, 61, 148 and 204 of the studies, respectively. Furthermore, study characteristics like the median year of recruitment, the number of patients for the follow-up and the average age of the patients (in 260 studies) were reported. Other variables provided information about the medication and medical history of the patients. For more information, we refer to \cite{kimmoun2021}; the authors made their dataset publicly available. 

\begin{figure}[b!]
\centering
\includegraphics[scale = 0.55]{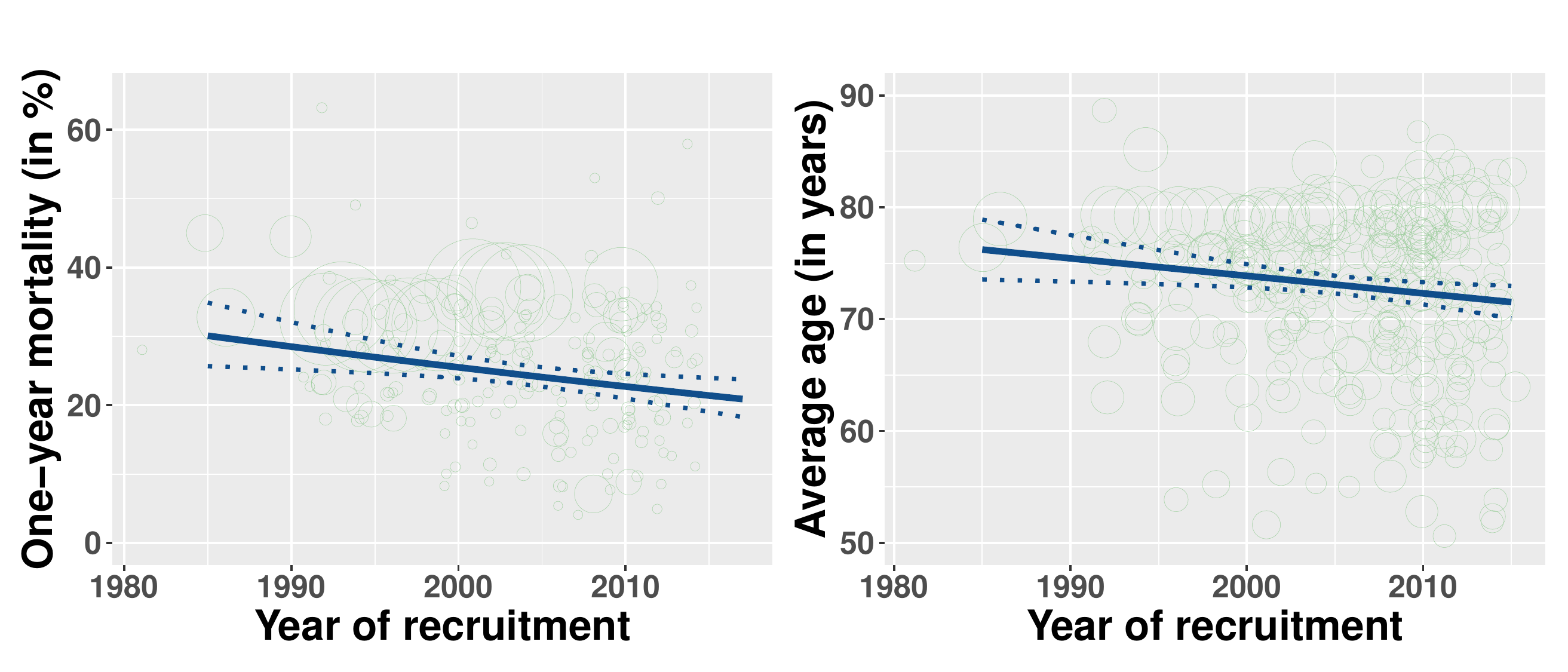}
\caption{Two models from the analysis by \cite{kimmoun2021}: Meta-regression model of year of recruitment for the one-year mortality (left) and weighted linear model of the year of recruitment for the mean age with the logarithmic study size as weights (right). Point sizes are proportional to the respective study sizes. Straight lines are the models' prediction, dotted lines the corresponding $95\%$ CI.}
\label{paper_plot}
\end{figure}

\subsection{Meta-regression analysis with an univariable model}
A major finding of the analysis of \cite{kimmoun2021} was a statistically significant decline in the $1$-year mortality over calendar time. The decline is shown in the left panel of Figure \ref{paper_plot} (adopted from Figures 2B \& 1A in
\cite{kimmoun2021}). A meta-regression with the logit transformed one-year mortality as the outcome variable $y_i$ from Equation (\ref{eq:om}) and the year of recruitment as the only explanatory variable $x_{1i}$ was conducted for $k=204$ studies. The estimates of $\beta_0$ and $\beta_1$ are $\hat{\beta}_0=29.432$ (95\%-CI: [7.328,  51.537]) and $\hat{\beta}_1=-0.015$ (95\%-CI: [-0.263, -0.042]). However, the average age is decreasing in the recruitment year as well. In the right plot of Figure \ref{paper_plot} it is shown that the average age of the participants decreased by $1.56$ years every ten years. Therefore, it is of interest how the effect of the recruitment year changes in a model where the mean age and their interaction are included as well.

\subsection{Meta-regression analysis with interaction}
\begin{figure}[t!]
    \centering
    \includegraphics[scale=0.55]{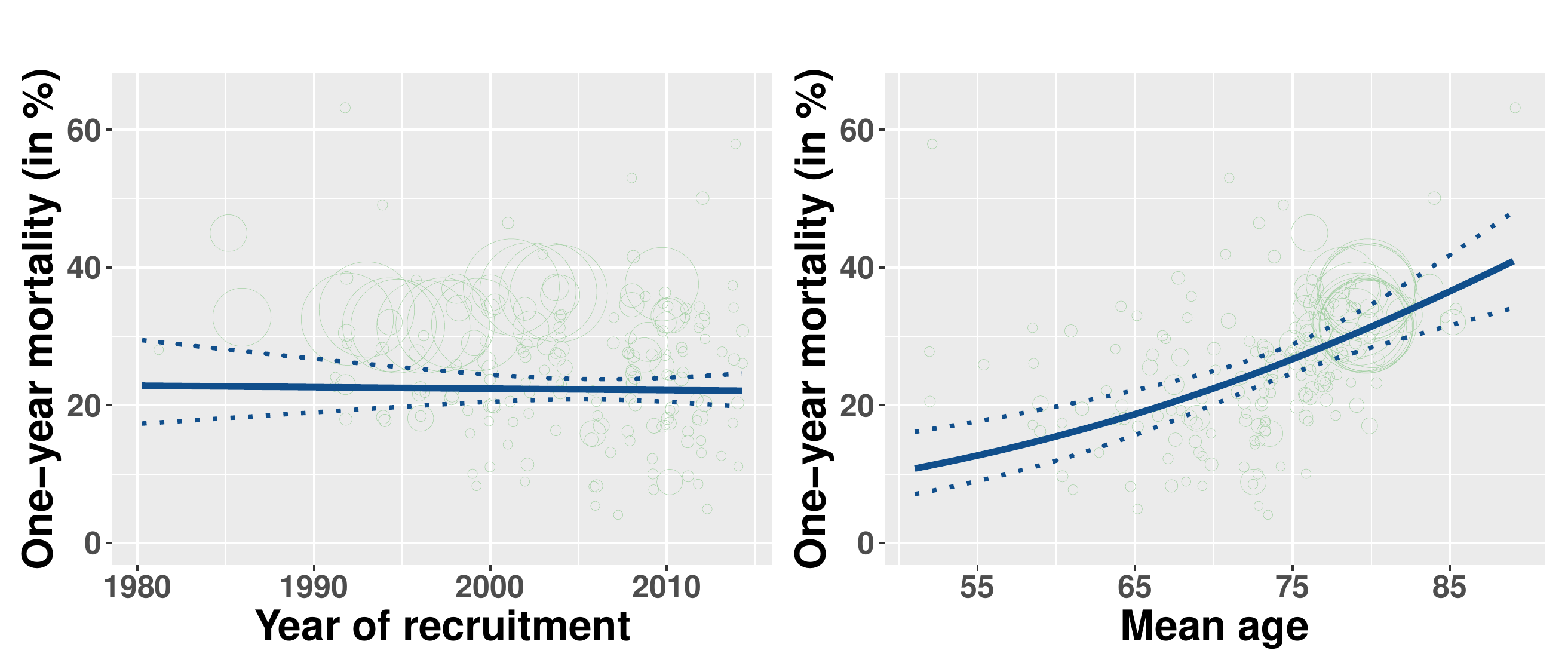}
    \caption{Trends of year of recruitment (left) and mean age (right) on the one-year mortality in a mixed-effects meta-regression model with these two moderators and their interaction. Point sizes are proportional to the respective study sizes. Straight lines are the models' prediction at the median level of the respective other explanatory variable, dotted lines the corresponding $95\%$ CI.}
    \label{new_example}
\end{figure}

For this purpose, we considered a model with two moderators and their interaction, where we used centred explanatory variables to reduce the impact of multicollinearity (\cite{aiken1991}, p.28-31). In the model the (logit transformed) one-year mortality was regressed on the year of recruitment ($x_{1i} $ in Equation (\ref{eq:om})), the average age ($x_{2i}$) and their interaction for $181$ studies. While the mean age has a significant effect on the one-year mortality ($\hat{\beta}_2=0.0333$, $95\%$-CI: $[0.0208, 0.0457]$) the year of recruitment turned out to be not significant ($\hat{\beta}_1=-0.0066$, $95\%$-CI: $[-0.0185, 0.0052]$) in the meta-regression model with two moderators and their interaction. The predicted effect of each explanatory variable at the median level of the respective other variable is shown in Figure~\ref{new_example}. 
The interaction term ($\hat{\beta}_{12}=-0.0018$, 95\%-CI: $[-0.0035, -0.0001]$) and intercept ($\hat{\beta}_{0}=-1.1477$, 95\%-CI: $[-1.2271, -1.0684]$) are significant at level $0.05$ as well. Hence, fitting a mixed-effects meta-regression model with interaction reveals that the apparent time trend is essentially captured by the changes in age over calendar time. This can already be seen in Figure~\ref{new_example} and is contrary to the conclusion drawn by \cite{kimmoun2021}. Their observed time trend seems to stem from the neglected correlation between the average age and the recruitment year, which was not considered in their meta-regression model with one moderator only.

\section{Simulation}\label{sec:sim}
To illustrate how strongly the neglect of an interaction term impacts the model, we conducted a small simulation study mimicking the setting in \cite{kimmoun2021}. We assumed that the estimated parameters in the model with interaction are the true parameters, i.e. we assumed $\boldsymbol{\beta}=(-1.1477, -0.0066, 0.0333, -0.0018)^T$ for the regression coefficients, $k=181$ for the number of studies and $\tau^2=0.2484$. The study sizes $n_i, i=1,...,k$, the centred recruitment year, average age and their interaction of study $i$ were assumed to be equal to the data example as well. Via the equation $\theta_i=\beta_0+x_{1i}\beta_1+x_{2i}\beta_2 + x_{1i}x_{2i}\beta_{12} + u_i$ we generated the true logit transformed mortality $\theta_i$, where the $u_i$ were drawn from a normal distribution with mean 0 and variance $\tau^2$. Hence, the probability for death within one-year in study $i$ was assumed to be $p_i=\exp(\theta_i)/(1+\exp(\theta_i))$. The number of deaths $d_i$ in study $i$ was sampled from a binomial distribution with parameters $n_i$ and $p_i$. Moreover, the observed logit transformed mortality $y_i$ was set to \[y_i=\log\left(\frac{d_i+0.5}{n_i-d_i+0.5}\right),\] and its sampling variance to $v_i=(d_i+0.5)^{-1}+(n_i-d_i+0.5)^{-1}$, see \cite{sutton2000methods}, p. 18. Note, that $d_i/(n_i-d_i)$ is equivalent to the ratio of the rates. The 0.5 was added in both equations to avoid dividing by 0, see \cite{viechtbauer2007}.

\subsection{Impact of neglected moderators and interaction}\label{ssec:neg_int}
Based on this procedure, we generated 10,000 data sets and fitted them by (i) a model with two moderators and their interaction, (ii) a model with two moderators and no interaction and (iii) a model with only one moderator. For all models, we calculated 95$\%$ confidence intervals of the estimated coefficients as well. The respective empirical coverages, median interval lengths and biases are shown in Table \ref{tab:bsp_sim}. The model with only one moderator has low coverage for both parameters, especially $\beta_1$ where the true parameter is covered in only $77.2\%$ of all cases. For the model with two moderators, the coverages of $\beta_0$ with 89.8\% and $\beta_2$ with 90.6\% are below the nominal confidence level as well. In the model with interaction, all coefficients are covered in approximately $95\%$ of all cases. For $\beta_0$ and $\beta_1$ the model with one moderator has the longest intervals. The lengths of the intervals from the model with two moderators tend to be slightly shorter than in the model with interaction. Moreover, the absolute biases are higher for the models without interaction. For $\beta_1$ the bias in the model with two moderators is almost three times as high in the absolute value as in the model with interaction. The bias in the model with only one moderator is again more than five times as high as the bias in the model with two moderators.

\begin{table}[b!]
    \centering
     \caption{Coverages (nominal level 0.95), median interval lengths and biases of different models when the model with interaction is the true model}
    \begin{tabular}{lr|r|r|r}
           &moderator/model  &  one & two & two and interaction\\\hline
\multirow{4}{*}{Coverage} & $\beta_0$ &  0.9159 &0.8979 & 0.9447          \\
                   & $\beta_1$ & 0.7718&  0.9547& 0.9495 \\
                   & $\beta_{2}$ & -  & 0.9062& 0.9527 \\
                   & $\beta_{12}$ &- &- & 0.9502\\
                  \hline
\multirow{4}{*}{Interval length}
& $\beta_{0}$ &  0.1615  & 0.1517 &   0.1535                \\
                  &  $\beta_{1}$ &0.0240 &0.0231 &   0.0230         \\
                  &  $\beta_{2}$ & - & 0.0239& 0.0242\\
                  &  $\beta_{12}$ & - &- &0.0034    \\
                 \hline
\multirow{4}{*}{Bias} 
                  & $\beta_{0}$ &0.0282 & 0.0264&0.0077 \\
                   & $\beta_{1}$ &-0.0078 &-0.0012 &0.0001\\
                  & $\beta_{2}$ &- &-0.0038 &0.0000 \\
                   & $\beta_{12}$ & -&- & -0.00004 \\\hline
\end{tabular}
    \label{tab:bsp_sim}
\end{table}

\begin{table}[t!]
\centering 
    \caption{Coverages (nominal level 0.95), median interval lengths and biases of different models when the model with only one moderator is the true model}
    \begin{tabular}{lr|r|r|r}
           &moderator/model  &  one & two & two and interaction\\\hline
\multirow{4}{*}{Coverage} & $\beta_0$ &  0.9431 &0.9495 & 0.9504        \\
                   & $\beta_1$ & 0.9495& 0.9481&  0.9501 \\
                   & $\beta_{2}$ & -  & 0.9544& 0.9539 \\
                   & $\beta_{12}$ &- &- & 0.9518\\
                  \hline
\multirow{4}{*}{Interval length}
& $\beta_{0}$ &  0.1619  & 0.1618 &   0.1663                \\
                  &  $\beta_{1}$ &0.0240 &0.0247 &   0.0249         \\
                  &  $\beta_{2}$ & - & 0.0253& 0.0262\\
                  &  $\beta_{12}$ & - &- &0.0037   \\
                 \hline
\multirow{4}{*}{Bias} 
                  & $\beta_{0}$ &0.0087 & 0.0092&0.0071 \\
                   & $\beta_{1}$ &0.0001 &0.0003 &0.0004\\
                  & $\beta_{2}$ &- &-0.0002 &-0.0003 \\
                   & $\beta_{12}$ & -&- & 0.0000 \\\hline
\end{tabular}
    \label{tab:bsp_sim2}
\end{table}

\subsection{Impact of redundant moderators and interaction}
The above results show the importance of including potential moderator and interaction effects when they are present. 
In the next step, we investigate possible drawbacks of more complex modelling in situations where only a single moderator effect is present. To this end we generated 10,000 univariate models and estimated them via a model with one moderator, two moderators or two moderators and their interaction. In order to make this simulation comparable to the one conducted in Section \ref{ssec:neg_int}, we used the centered moderators and only the 181 studies that were included in the model with interaction to obtain parameter estimates. Thus, we set $\tau^2=0.2935$ and $\boldsymbol{\beta}=(-1.1497, -0.0148, 0, 0)^T$, i.e. we only have a model with intercept and one moderator. The empirical coverages, median interval lengths and biases of all estimated models are shown in Table \ref{tab:bsp_sim2}. Here, the coverages in the model with interaction are only slightly lower than in the true model with only one moderator. For $\beta_0$ and $\beta_1$ the median interval length is slightly larger in the model with interaction, compared to the model with only one moderator. The bias of $\beta_0$ is also slightly larger in the model with interaction. Regarding the bias of $\beta_1$ the two models do not differ.

\section{Discussion}
Motivated by a meta-regression on a temporal trend in mortality of acute heart failure patients \citep{kimmoun2021}, we considered the impact of omitted and unnecessarily fitted interactions in meta-regression. Reanalysing the study of \cite{kimmoun2021}, we showed that the time trend found in the study seems to be caused by a neglected interaction with the age of the patients. By means of a small simulation study, we showed that fitting a more complex model is related to longer confidence intervals, whereas neglecting an interaction may cause a large bias in the estimation and too liberal coverages. This leads us to the recommendation to always include an interaction to the model when in doubt. Otherwise, wrong conclusions may be drawn.\\ The presented simulation is motivated by the example considered and therefore limited in scope. Nevertheless, it seems plausible that the same conclusions can be drawn for models of different specification, since the effects of misspecification reflect the ones known from ordinary least squares regression. To verify this research hypothesis, an extended simulation study with varying parameters and model specification is necessary and will be conducted in the future.  


\subsection*{Financial disclosure}

 This work was supported by the German Research Foundation projects (Grant no. FR 3070/3-1 (Tim Friede) and PA-2409 7-1 (Markus Pauly)).

\subsection*{Conflict of interest}

The authors declare no potential conflicts of interest.

\subsection*{Data availability statement}
The R-script used for the analysis of the data example and the simulation study will be made publicly available on osf.io under https://osf.io/t5zsc/. The dataset from \cite{kimmoun2021} was made publicly available by the authors themselves under https://osf.io/cxv5k/.

\section*{Highlights}
What we know so far: 
\begin{itemize}
    \item In Meta-regression interaction terms of moderators are often not considered.
    \item Omitting important variables in a regression model can lead to biased estimates, whereas including redundant variables increases the variance of estimates.
\end{itemize}
What is new:
\begin{itemize}
    \item By reanalyzing a meta-analysis on acute heart failure, the study illustrates how neglecting an interaction can alter the results in a meta-regression model.
    \item A small simulation study is conducted, which shows how neglecting and unnecessarily including moderators and interaction effects confidence intervals estimates in meta-regression.
\end{itemize}
Potential impact for RSM readers outside the authors’ field:
\begin{itemize}
    \item The results suggest to always include interaction terms in meta-regression, when such interactions are plausible, because otherwise confidence intervals of parameter estimates might have poor coverage.
\end{itemize}

\bibliographystyle{plainnat}
\bibliography{Literatur}

\end{document}